\documentclass[preprint,12pt]{elsarticle}

\makeatletter
\def\ps@pprintTitle{%
 \let\@oddhead\@empty
 \let\@evenhead\@empty
 \def\@oddfoot{\centerline{\thepage}}%
 \let\@evenfoot\@oddfoot}
\makeatother

\newcommand{\usepackageifexists}[1]{%
     \IfFileExists{#1.sty}{\usepackage{#1}}%
        {\GenericInfo{taglia}{Il package #1 non esiste.}}}

\makeatletter
\def\paragraph{\@startsection{paragraph}{4}%
  \z@\z@{-\fontdimen2\font}%
  {\normalfont\bfseries}}
\makeatother

\usepackage{verbatim}
\usepackage{amsmath,amsthm,amssymb}
\usepackage{amsthm, bm}
\theoremstyle{plain}

\theoremstyle{definition}

\theoremstyle{remark}
\newtheorem{remark}{Remark}

\usepackage{upref}
\usepackage{graphicx, caption, subcaption}

\usepackage{listings}
\usepackage{epstopdf}

\usepackage{color}
\usepackage{mathrsfs}
\usepackage{multirow}
\providecommand{\mathscr}[1]{\mathcal{#1}}

\usepackage{graphicx}
\usepackageifexists{pdfsync}
\graphicspath{ {figures/} }
\usepackage{booktabs}
\usepackage{hyperref}
\usepackage{geometry, natbib} 
\usepackage{adjustbox}
\usepackage[autostyle]{csquotes} 
\usepackage[para,online,flushleft]{threeparttable}

\newcommand{\PP}{\mathbb{P}}
\newcommand{\F}{\mathcal{F}}

\def\correspondingauthor{\footnote{Corresponding author.}}

\begin{document}

\title{Forecasting interest rates through Vasicek and CIR models: a partitioning approach}

\author[Orlando]{Giuseppe Orlando\correspondingauthor{}}
\address[Orlando]{Universit\`a degli Studi di Bari "Aldo Moro" - Department of Economics and Mathematical Methods, Via C. Rosalba 53, Bari, I-70124 Italy, Tel. +39 080 5049218, giuseppe.orlando@uniba.it}

\author[Mininni]{Rosa Maria Mininni} 
\address[Mininni]{Universit\`a degli Studi di Bari "Aldo Moro" - Department of Mathematics, Via Orabona 4, Bari, I-70125 Italy, Tel. +39 080 544 2700, rosamaria.mininni@uniba.it}
 
\author[Bufalo]{Michele Bufalo} 
\address[Bufalo]{Universit\`a degli Studi di Roma "La Sapienza" - Department of Methods and Models for Economics, Territory and Finance, Via del Castro Laurenziano 9, Roma, I-00185, Tel. +39 06 49766903} 

\begin{abstract}
The aim of this paper is to propose a new methodology that allows forecasting,
through Vasicek and CIR models, of future expected interest rates (for each maturity) based on rolling windows from observed financial market data. The novelty, apart from the use of those models not for pricing but for forecasting the expected rates at a given maturity, consists in an appropriate partitioning of the data sample. This allows capturing all the statistically significant time changes in volatility of interest rates, thus giving an account of jumps in market dynamics. The performance of the new approach is carried out for different term structures and is tested for both models. It is shown how the proposed methodology overcomes both the usual challenges (e.g. simulating regime switching, volatility clustering, skewed tails, etc.) as well as the
new ones added by the current market environment characterized by low to negative interest rates.
\end{abstract}


\maketitle

\section{Introduction}\label{sec:Introduction}

The present paper has the objective of forecasting interest rates (by maturity)
from observed financial market data through a new approach that preserves the analytical tractability of the stochastic models describing the dynamics of real market interest rates proposed by Vasicek (Vasicek, 1977) and Cox-Ingersoll-Ross (CIR) (Cox, Ingersoll \& Ross, 1985). This because of their popularity within the financial community given their simplicity (uni-factorial, mean reverting models) and their ability to provide closed form solutions for pricing interest rate derivatives (Zeytun \& Gupta, 2007). The idea of this work is to overcome both the usual challenges imposed by regime switching, volatility clustering, skewed tails, etc., as well as the new ones added by
the current market environment (particularly the need to model a downward trend to negative interest rates). This is to be achieved by proposing a new methodology that allows forecasting of future expected interest rates by an appropriate partition of the dataset and assuming that the dynamic of each rate is represented by the Vasicek or CIR model. The effect of partitioning the available market data into sub-samples with an appropriately chosen probability distribution is twofold: (1) to improve the calibration of the Vasicek/CIR model's parameters in order to capture all the statistically significant changes of variance in market spot rates and so, to give an account of jumps; (2) to consider only the most relevant historical period. The distributions herein considered for the dataset partition are the Normal and noncentral
Chi-square distribution. These distributions have been chosen by analogy
with the steady (resp. conditional) distribution of the interest rate process in the Vasicek (resp. CIR) model. The performance of the new approach, tested on weekly EUR data on bonds with different maturities, has been carried out for both Vasicek and CIR model, and compared with the Exponentially Weighted Moving Average (EWMA) model in terms of forecasting error. The error analysis highlighted a better performance of the proposed procedure with respect to the EWMA and better results in prediction when a partition with non-central Chi-square distribution (CIR model) is considered.
 
This paper is organized as follows: Section~\ref{sec:Literature} illustrates the rationale behind both the Vasicek and the CIR models and summarizes the existing literature.  Section~\ref{sec:Model_Data} explains why to implement a new methodology. Section~\ref{sec:Model} presents the model in full detail.  Section~\ref{sec:EmpResults} shows the empirical results for weekly recorded EUR interest rates in both money market and short to long-term datasets. Section~\ref{sec:Conclusions} contains the conclusions.

\section{Background and Literature Review}\label{sec:Literature}

Usually, the dynamics of interest rates is mathematically described by a stochastic differential equation (SDE) of type
\begin{equation}\label{SDEr}
dr(t)=\mu(t,r(t))dt+\sigma(t,r(t))dW(t), \quad r(0) = r_0,
\end{equation}
with drift $\mu(\cdot,\cdot)$, diffusion term  $\sigma(\cdot,\cdot)$ and $(W(t))_{t\ge 0}$ a standard Brownian motion. The unique solution to \label{SDEr}, say $r=(r(t))_{t\ge 0}$, in case it exists, is a diffusion process, i.e. a continuous Markov process defined on a given probability space $(\Omega, \F, \PP)$.  

The Vasicek and the CIR model in \eqref{SDEr} have the form
\begin{equation}\label{SDEVC}
dr(t)= \kappa (\theta -r(t)) dt + \sigma(t,r(t)) dW(t), \quad r(0) = r_0,
\end{equation}
where the constant parameters $\kappa $ and $\theta $  denote the reversion's speed and the long-term mean, respectively. The rationale behind the mean reversion is that higher rates slow down the economy and reduces the demand for funds. The opposite when rates are low. However this has proved to be not always true as ``you can take a horse to water, but you can't get it to drink".

For the Vasicek model, $\kappa,\, \theta>0,\,$ and the volatility $ \sigma(t,r(t))$ is a constant parameter $\sigma >0$.  The process $r$ is known as the  {\it Orstein-Uhlenbeck} process. Further, $r$ has a steady normal distribution with mean $\theta$  and long-term variance $\sigma ^{2}/2 \kappa$. This allows the positive probability of getting negative interest rates, which was not expected before the massive injection of liquidity and credit facilities provided by central banks following 2008 credit squeeze. Among the drawbacks of this model are the poor fitting to the current term structure of interest rates (later addresses by Hull \& White, 1990) and the undesired property that the yields over all maturities are perfectly correlated. Moreover, the conditional volatility of changes in the interest rate is constant, independent on the level of r, which can unrealistically affect the prices of bonds (see Rogers, 1996).

For the CIR model, $\kappa,\, \theta > 0 $ and the volatility $ \sigma(t,r(t)) = \sigma \sqrt{r(t)} >0$ in \eqref{SDEr}, with $\theta\sigma ^{2}/2 \kappa$ the long-term variance. The process $r$ is known as the {\it square-root process}. The conditional distribution of $r$ is a non-central Chi-square distribution and the steady distribution is a Gamma. Therefore, the square-root  process $r$  is always non-negative; it is known that if  the involved parameters satisfy the condition $2k\theta > \sigma ^{2},$ then $r(t)$ is strictly positive for any $t\ge 0$, and, for small $r(t)$, the process rebounds as the random perturbation dampens with $r(t)\to 0$. 
The relatively handy implementation and tractability of the CIR model, as well as the specific characteristic of precluding negative interest rates, an undesirable feature under 2008 pre-crisis assumptions, are two reasons that allowed the CIR model to become one of the most widely used short-term structure models in finance. However, there are a number of issues in describing interest rate dynamics within the CIR framework such as: 
\phantomsection \label{CIR-Issues}

\noindent {\bf a)} interest rates can never reach negative values; {\bf b)} the diffusion term $\sigma\sqrt{r(t)}$  goes to zero when $r(t)$ is small, in contrast to actual experience; {\bf c)} the volatility parameter $\sigma$ is constant, whereas in reality $\sigma$  changes continuously; {\bf d)} as well as with the Vasicek model, there are no jumps neglecting in this way events such as fiscal and monetary decisions, release of corporate financial results, changes in investors' expectations, etc. 

In fact, the current market environment with abrupt changes in volatility and negative interest rates has exacerbated the above mentioned issues urging the need for more sophisticated frameworks, which could accommodate multiple sources of risks, as well as shocks and/or structural changes of the market. This has lead to the development of a number of papers for pricing interest rate derivatives that are based on stochastic interest rate models generalizing the classical CIR and Vasicek paradigm (for more details the reader can refer to Brigo \& Mercurio, 2006, Ch. 3-4, and references therein). More recently, Zhu (Zhu, 2014) proposed a CIR variant with jumps modelled by a Hawkes process, and Moreno and Platania (Moreno \& Platania, 2015) presented a cyclical square-root model, where the long-run mean and the volatility parameters are driven by harmonic oscillators. Finally Naja and Mehrdoust (Naja \&Mehrdoust, 2017) and Naja et al. (Naja, Mehrdoust, Shirinpour, \& Shima, 2017) proposed some extensions of the CIR framework where a mixed fractional Brownian motion is added to account for the random part of the model.


\section{Material and Method}\label{sec:Model_Data}

\subsection{Dataset}\label{sec:Dataset}
Our dataset records weekly (spanning from 31 December 2010 to 18 November 2016) EUR interest rates with maturities 1/360A, 30/360A, 60/360A,..., 360/360A and 1Y,..., 50Y (i.e. at 1 day (overnight), 30 days, 60 days,...., 360 days  and  1 Year,...,50 Years) available from IBA \footnote{ICE Benchmark Administration, Data Vendor Codes.} \cite{IBA}. For our convenience we have split the data in two Datasets: money market (Dataset I) and short- to long-term interest rates (Dataset II).  
%

\begin{table}[!ht] 
	\centering
	\caption{Weekly EUR interest Rates: the Dataset}
	\label{tab:DatasetW}
	\begin{tabular}{c|cccc|cccc} 
		\hline\noalign{\smallskip}
		&\multicolumn{4}{c|}{\bf Dataset I}&\multicolumn{4}{c}{\bf Dataset II}\\ 
		\noalign{\smallskip}\hline\noalign{\smallskip}
		&\multicolumn{8}{c}{\bf Maturity}\smallskip\\
		\bf Date&\bf 1/360A&\bf 30/360A&$\cdots$&\bf 360/360A&\bf 1Y&\bf 2Y&$\cdots$&\bf 50Y\\ 
		\noalign{\smallskip}\hline\noalign{\smallskip}
		31.12.2010&0.606 & 0.782 &$\cdots$ & 1.507&1.311&1.557&$\cdots$&3.306\\
		07.01.2011&0.341&0.759&$\cdots$&1.505&1.345&1.603&$\cdots$&3.229\\
		\vdots&\vdots&\vdots&$\cdots$&\vdots&\vdots&\vdots&$\cdots$&\vdots\\
		18.11.2016&-0.410&-0.373&$\cdots$&-0.077&-0.195&-0.139&$\cdots$&1.120\\ 
		\noalign{\smallskip}\hline
	\end{tabular}
\end{table}

\begin{figure}[!htbp]
	\centering
	\includegraphics[width=\textwidth]{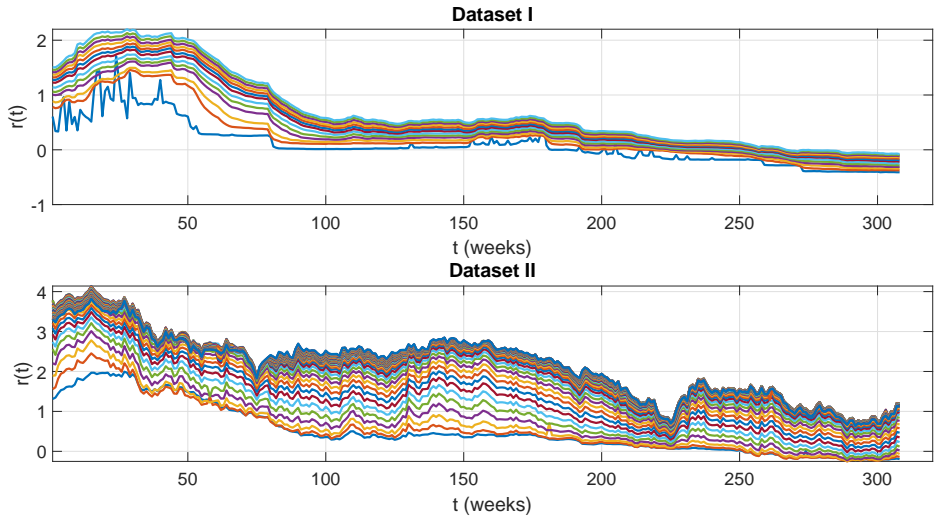}
	\caption{Datasets I and II: weakly observed EUR interest rates}
	\label{Fig:InterestRatesW}
\end{figure}

In Table~\ref{tab:DatasetW}),  each column lists a sample of  $n=308$ weakly observed EUR interest rates with a set maturity; each row shows interest rates on different maturities observed at a fixed date. 

The plots in Figure~\ref{Fig:InterestRatesW} represent the columns of Dataset I and II, so they are different from the yield curves (term structure) by plotting the rows. From Dataset I it is evident that  the short-term rates become permanently negative after 2014 (as from March 2015). However, sample data from Dataset II also show a downward trend. 

In (Orlando, Mininni, \& Bufalo, 2018b) we carried out a qualitative analysis of a dataset based on monthly observed spot rates and showed that the most challenging task is to fit money market interest rates, due to the largest presence of close-to-zero and/or negative spot rate values. For this reason, we start to examine samples of interest rates with maturity from Dataset II. In this paper we limit ourselves to estimate and forecast future expected interest rates over rolling time windows of market data. A detailed analysis of Dataset I linked to the problem of estimating and forecasting  interest rates will be treated in an outgoing research.

\section{The Model: procedure and accuracy}\label{sec:Model}

Financial time series for interest rates frequently show an empirical distribution as a mixture of probability distributions with sudden changes in the magnitude of variations of the observed values. Thus, in order to capture all the statistically significant changes of variance in market spot rates and so, to give an account of jumps, the available market data sample is partitioned into sub-samples - not necessarily of the same size - with a Normal or a non-central Chi-square distribution by using an appropriate technique described in Section~\ref{sec:Partition}. 
Further, in case negative or close-to-zero market interest rates are present in the observed dataset, the procedure involves a shift to positive values by a suitable scalar parameter (Section~\ref{sec:IRShift}). This to avoid that the diffusion term in \eqref{SDEVC} is not dampened by the proximity to zero, as for the CIR model, but fully reflects the same level of volatility present on the market .

Finally, to deal with the crucial issue of a constant volatility parameter $\sigma$ and the problem of an unsatisfactory calibration to market data for both CIR and Vasicek models, we  calibrate, for each sub-sample, the model's parameters to the observed interest rates (Section~\ref{sec:Calibration}). 

The proposed procedure is first tested on some  samples from the available dataset (Section~\ref{sec:EmpResults}) and then is applied to predict future expected next-week interest rates based on rolling windows, as described in Section~\ref{sec:Forecast}. 

\subsection{Step 1- Dataset Partition}\label{sec:Partition}

As observed in the Introduction, the novelty in our procedure consists in partioning the available market data into sub-samples with an appropriately chosen probability distribution. The effect of partitioning is twofold: (1) to improve the calibration of the Vasicek/CIR model's parameters to take account of multiple jumps in the dynamics of market interest rates and so, of time changes in spot rates volatility;  (2) to determine only the latest historical period over which predicting expected future interest rates that closely follow market rates.   The sub-samples are chosen according to the data empirical probability distribution, which is unknown. 

\noindent Notice that in the literature there are several approaches for detecting multiple changes  in the probability distribution of a stochastic process or a time series (see, for instance, for instance, Lavielle, 2005; Lavielle \& Teyssiere, 2006; Bai \& Perron 2003).
We shall use these methodologies in an ongoing research work, but in this paper we adopt the numerical partition into sub-samples following a Normal or a non-central Chi-square distribution, as described in the next subsections.

\subsubsection{Partition with Normal Distribution}\label{Sec:PartitionNorm}

In  (Orlando, Mininni \& Bufalo, 2018a) we hypothesized the empirical distribution of the observed data sample to be a mixture of normal distributions for the presence of negative interest rate values. 
This hypothesis is appropriate because in the Vasicek model the interest rate process $r$ has a steady normal distribution, as already observed in Section \ref{sec:Literature}. Moreover, the dynamics of the form \eqref{SDEVC} for the square-root process in the CIR model is obtained from a squared Gaussian model (see, for example, Rogers, 1996). Our idea was, therefore, to divide the data sample into a number of sub-samples each coming from  an appropriate normal distribution. The goodness-of-fit to a normal distribution is checked with the Lilliefors test at a $5\%$ significance level. This as an improvement on the Kolomogorov-Smirnov test where the population mean and standard deviation are not known, but are estimated from data. 
In this paper we have implemented a \emph {forward} procedure that starts by considering the first four data of the original sample, say $(r_1,\dots,r_4),$ and performs the Lilliefors test until the first normally distributed sub-sample, say $(r_1,\dots, r_{n_1}),$ with $n_1\ge 4,$ is not rejected. Then, the procedure is applied to the remaining sequence  $(r_{n_1+1},\dots, r_{n_1+4})$ until  the second normally distributed sub-sample $(r_{n_1+1},\dots, r_{n_2}),$ with $n_2\ge n_1+4,$ is not rejected and so on, up to partition the entire data sample into $m$ normally distributed sub-samples, namely $(r_1,..., r_{n_1}),$  $(r_{n_1+1},...,r_{n_2}),\dots,(r_{n_{m-1}+1},...,r_{n_m}),\, n_m\le n.$
Table \ref{tab:ForwardProcedure} summarizes the  \emph {forward} segmentation procedure.

\begin{table}[!h]\small
	\caption{“Forward” procedure}
	\label{tab:ForwardProcedure}
	\centering
	\begin{threeparttable}
		\begin{tabular}{|l|}
			\hline\noalign{\smallskip}
			1. Initialize h=4;\\
			2. run the Lilliefors test on the interest rate vector r(1:h);\\
			3. \textbf{while} the null hypothesis is not rejected \\
			4. h=h+1;\\
			5. run the Lilliefors test on r(1:h);\\
			6. \textbf{end}\\
			7. set n(1)=h;\\
			8. initialize i=1;\\
			9. \textbf{while} n(i)$<$length(r)\\
			10. h=n(i)+4;\\
			10. repeat steps 2-6 for r(n(i)+1:h) and find n(i+1);\\
			11. \textbf{if} length(r)-n(i+1)$<$4\\
			12. set resti=r(n(i+1)+1:length(r));\\
			13. \textbf{break}\\
			14. \textbf{else}\\
			15. set i=i+1;\\
			16. \textbf{end}\\
			17. \textbf{end}\\
			\noalign{\smallskip}\hline
		\end{tabular}
	\end{threeparttable}
\end{table}

Further, observe that in performing the Lilliefors test it could happen that the p-value is greater than the chosen significance level, but differs from it by no more than $10^{-2}$. Thus, in this case,  the Johnson transformation  is applied to ensure that each sub-sample follows a normal distribution.
The Johnson's method consists in transforming a non-normal random variable $X$ to a standard normal variable $Z$ as follows,
\begin{equation}\label{Jtransform}
Z=\gamma +\delta f\left(\frac{X-\xi}{\lambda}\right),\quad \lambda,\; \delta>0
\end{equation}
where $f$ must be a monotonic function of $X$ with the same range of values of the standardized random variable $(X-\xi)/\lambda,$ where $\xi $ and $\lambda $ are respectively the mean and the standard deviation of $X.$ The parameters $\delta$ and $\gamma$ reflect respectively the skewness and kurtosis of $f.$ The algorithm to estimate the four parameters $\gamma$, $\delta$, $\lambda$ and $\xi$, and to perform the appropriate transformation is available as a Matlab Toolbox written by Jones (Jones, 2014).

To apply the Johnson's method to our case, the market interest rates in each sub-sample have been first transformed by \eqref{Jtransform} to $m$ sub-samples with standard normal distribution that is, for any $ k=1,\dots,m,$ 
$$   
z_h=\gamma +\delta f\left(\frac{r_h-\mu_{k}}{\sigma_{k}}\right), \quad h=n_{k-1}+1,..,n_k\; (n_0=0),
$$
where $\mu_{k},\, \sigma_{k}$ denote respectively the sample mean and standard deviation of the $k$-th sub-group.  Then, they are transformed to $m$ sub-samples with normal distribution $N(\mu_{k},\sigma_{k})$ as follows
$$
r_h=\sigma_{k}z_h+\mu_{k},\quad h=n_{k-1}+1,..,n_k\; (n_0=0).
$$ 

\subsubsection{Partition with non-central Chi-square Distribution}\label{Sec:PartitionNCS}
As an alternative to the previous hypothesis of a mixture of normal distributions,  the empirical distribution of the analysed data sample may be assumed a mixture of non-central Chi-square distributions. This hypothesis is justified from the conditional distribution of the CIR process, as mentioned in Section  \ref{sec:Literature}. Since the non-central Chi-square distribution admits only positive values, 
the market observed interest rates have to be first shifted to positive values as described in the next subsection. The partitioning procedure is analogous to that described in Section~ \ref{Sec:PartitionNorm} and  the Kolmogorov-Smirnov test is performed (at a $5\%$ significance level) to test the goodness-of-fit of the $m$ sub-samples to a non-central Chi-square distribution.

\subsection{Step 2 - Interest Rates Shift}\label{sec:IRShift}

As mentioned above, a step of the procedure consists in translating market interest rates to positive values to eliminate negative/near-zero values and not to dampen the volatility in the CIR model. 
Herein we consider the following transformation
\begin{equation}\label{eq3.21}
r_{shift}(t) = r(t) + \alpha,\quad t\in [0,T],
\end{equation}
where $\alpha$ is a deterministic positive quantity. This translation leaves unchanged the stochastic dynamics \label{CIRrate} of the interest rates i.e., for any time $t,$ $dr_{shift}(t)=dr(t).$ There are many values that could be assigned to $\alpha$, but we believe that the most appropriate choice is the $99th$ percentile of the empirical interest rates probability distribution. If the translation \eqref{eq3.21} is not adequate to move negative interest rates to corresponding positive values, which means further negative values are  between the $99th$- and the $100th$-percentile, we can set $\alpha$ equal to the $1st$-percentile of the empirical distribution. In this case \eqref{eq3.21} becomes 
$$
r_{shift}(t)=r(t)-{\alpha}.
$$

\subsection{Step 3 - Calibration}\label{sec:Calibration}
In order to estimate interest rates from the Vasicek and CIR model, the involved parameters $k, \theta, \sigma$ need to be calibrated to the market interest rates. 
In the present work, among many approaches existing in the literature to estimate the parameters of a SDE (see, for instance, Poletti Laurini \& Hotta, 2017, and references therein), we
applied the estimating function approach for ergodic diffusion models introduced in
Bibby et al. (Bibby, Jacobsen, \& S\o{}rensens, 2010), which turned out to be very useful in obtaining optimal estimators for the parameters of discretely-sampled ergodic Markov processes whose likelihood function is usually not explicitly known. In (Orlando, Mininni \& Bufalo, 2018a; 2018b) a better performance of the latter method is shown by comparing its efficiency with the maximum likelihood estimation routine implemented in Matlab for the CIR process by Klad\'{\i}vko (Klad\'{\i}vko, 2007). 
In (Bibby, Jacobsen, \& S\o{}rensens, 2010, Example 5.4) the authors constructed an approximately optimal estimating function for the CIR model,  from which they derived the following explicit estimators of the three parameters $\kappa, \theta, \sigma$ based on a sample of $n$ observed market spot rates $(r_1,\dots,r_n)$: 
\begin{align}\label{parCIR}
\hat\kappa_n &=-\ln\left(\frac{(n-1)\sum_{i=2}^n r_i/r_{i-1} - (\sum_{i=2}^n r_{i}) (\sum_{i=2}^n r^{-1}_{i-1})}{(n-1)^{2}-(\sum_{i=2}^n r_{i-1})(\sum_{i=2}^n r^{-1}_{i-1})}\right),\nonumber\\
\hat {\theta}_n &=\frac{1}{(n-1)}\sum_{i=2}^n r_{i}+\frac{e^{-\hat\kappa_n}}{(n-1)(1-e^{-\hat\kappa_n})}(r_{n}-r_{1}),\\[1.5\jot]
\hat {\sigma}^2_n &=\frac{\sum_{i=2}^n r^{-1}_{i-1}(r_{i}-r_{i-1}e^{-\hat\kappa_n}-\hat {\theta}_{n}(1-e^{-\hat\kappa_n}))^{2}}{\sum_{i=2}^n r^{-1}_{i-1}((\hat {\theta}_{n}/2-r_{i-1})e^{-2\hat\kappa_n}-(\hat {\theta}_{n}-r_{i-1})e^{-\hat\kappa_n}+\hat {\theta}_{n}/2)/\hat\kappa_{n}}.\nonumber
\end{align}
Similar calculations allow to compute in closed-form the estimators of the three parameters $\kappa, \theta, \sigma$ for the Vasicek model:
\begin{align}\label{parV}
\hat\kappa_n &=-\ln\left( \frac{\frac{1}{(n-1)}\bigl(\sum_{i=2}^n r_{i-1}\bigr)\bigl(\sum_{i=2}^n r_{i}\bigr)-\sum_{i=2}^n r_{i-1}r_i  }{\frac{1}{(n-1)}\bigl(\sum_{i=2}^n r_{i-1}\bigr)^2-\sum_{i=2}^n r^2_{i-1}   }\right),\nonumber\\
\hat {\theta}_n &= \frac{1}{(1-e^{-\hat\kappa_n})} \biggl( \frac{1}{(n-1)}\sum_{i=2}^n r_i-\frac{e^{-\hat\kappa_n}}{(n-1)}\sum_{i=2}^n r_{i-1} \biggr),\\[1.5\jot]
\hat {\sigma}^2_n &=\frac{2 \hat\kappa_n}{(1-e^{-2\hat\kappa_n})}\cdot \frac{1}{(n-1)}\sum_{i=2}^n \bigl( r_i-r_{i-1}e^{-\hat\kappa_n}-\hat \theta_n (1-e^{-\hat\kappa_n})\bigr)^2.\nonumber
\end{align}

\begin{remark}\label{Cal_remark}
Notice that the estimators given in \eqref{parCIR} and \eqref{parV} exist provided that the expression for $e^{-\hat\kappa_{n}}$ is strictly positive (Bibby, Jacobsen, \& S\o{}rensens observed that this happens with a probability tending to one as $n\to\infty$).

\noindent It is worth noting that in the presence of negative/near-zero interest rate values, as shown in Figure \ref{Fig:InterestRatesW}, the calibration of the unknown parameters for the CIR model may be carried out only after shifting spot rates to positive values by using the transformation \eqref{eq3.21}. 
\end{remark}

\subsection{Step 4 - Forecasting}\label{sec:Forecasting}
After calibration to the market data, the estimates of the parameter vector $(k, \theta, \sigma)$ have been used to forecast future expected interest rates 
by the following conditional expectation formula available in closed form for the interest rate process in both Vasicek and CIR model \eqref{SDEVC}
\begin{equation}\label{Expectation}
E[r(t) | r(s) ] = \theta + (r(s) - \theta) e^{-k(t-s)},\quad   0\le s<t.
\end{equation}

\subsection{Accuracy}\label{sec:Accuracy}
In order to measure the accuracy of  our approach, we compute the square-root of the mean square error (RMSE), say $ \varepsilon$, defined as 
\begin{equation}\label{sqerror}
\varepsilon= \sqrt{\frac1n\sum_{h=1}^{n} e_h^{2}},
\end{equation} 
where $e_h = r_{h} - \widehat{r}_h$ denotes the residual between the market interest rate $r_{h}$ and the corresponding fitted value $\widehat{r}_h.$ 
In our case the fitted values are the expected interest rates estimated through the numerical procedure described in the following subsections and compared to market data in Section~\ref{sec:EmpResults}. 

\section{Empirical Results}\label{sec:EmpResults}
\begin{figure}[!h]
	\centering
	\includegraphics[width=0.8\textwidth]{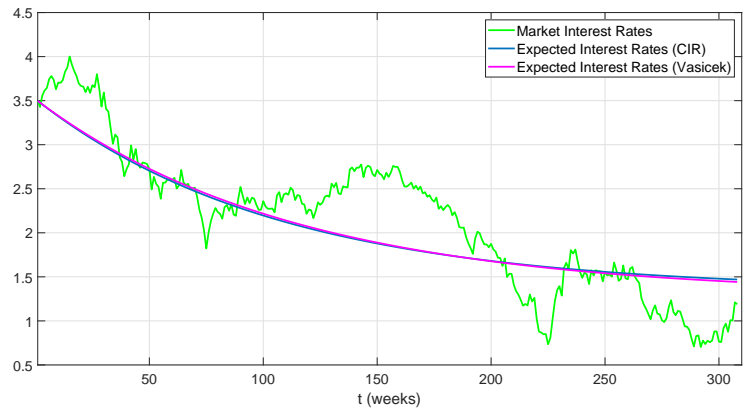}
	\caption{Estimated expected interest rates: CIR model (blue line) and Vasicek model (magenta line) versus $n=308$ weekly observed EUR interest rates (green line) with maturity $T=$ 30Y from Dataset II in Table \ref{tab:DatasetW}.}
	\label{fig:figure2}
\end{figure}

In order to test the performance of the methodology herein proposed, some empirical investigations have been done using two data samples of the dataset reported in Table \ref{tab:DatasetW}. 
As observed in Section~\ref{sec:Dataset}, we started to examine a market data sample from Dataset II. We considered a sample consisting of $n=308$ weekly observed market interest rates on derivatives with maturity $T=$ 30Y.  Figure~\ref{fig:figure2} shows that the observed interest rates are all positive with near-to-zero values in the tail (green line). 

We begin to estimate the expected interest rates from the Vasicek and CIR models. 
To calibrate the parameter vector $(k, \theta, \sigma)$ in both models to the market data, we applied the optimal estimating function method mentioned in Section~\ref{sec:Calibration}.  Notice that for the CIR model, we first shifted the whole data sample away from zero by formula \eqref{eq3.21}. Then the optimal parameter estimates have been used to calculate the estimated expected interest rates, say $(\widehat r_{exp}(t))_{t\geq 0}$, by formula \eqref{Expectation}.
The initial value has been set equal to the first value in the observed data sample.
Table~\ref{tab:Estimates} lists the  parameter estimates $(\hat{k}_n,\, \hat{\theta}_n,\, \hat{\sigma}_n)$  and the corresponding RMSE $\varepsilon$ for both models. In Figure \ref{fig:figure2} the original market data sample with the corresponding sequence of the estimated expected CIR/Vasicek interest rates are compared.
\begin{table}[!h] 
	\centering
	\caption{Optimal parameter estimates and the corresponding RMSE $\varepsilon$ for the Vasicek and CIR model.  Data sample: $n=68$ monthly observed interest rates with maturity $T=$ 30Y from Dataset II in Table \ref{tab:DatasetW}.}
	\begin{tabular}{ccc}
		\hline\noalign{\smallskip}
		&\multicolumn{2}{c}{\bf Parameter Estimates} \\
		\noalign{\smallskip}
		& {\bf Vasicek} & {\bf CIR} \\
		\noalign{\smallskip}\hline\noalign{\smallskip} 
		\bf $\hat\kappa_n$  &0.0087 &0.0094\\
		\bf $\hat {\theta}_n$ &5.1470 &5.2037\\
		\bf $\hat {\sigma}_n$ &0.0918 & 0.0379\\ 
		\noalign{\smallskip}\hline\noalign{\smallskip}		
		\bf $\varepsilon$  &0.4411 &0.4324\\ 
		\noalign{\smallskip}\hline
	\end{tabular}
	\label{tab:Estimates}
\end{table}

To improve  the results shown in Figure \ref{fig:figure2} in terms of fitting closely the market data, we implemented a numerical algorithm based on the following main steps summarizing the model described in Sections~\ref{sec:Partition}-\ref{sec:Calibration} :
\begin{enumerate}
	\item Shifting each sub-group to positive values by using the translation formula \eqref{eq3.21} (if needed);
	\item Partitioning the whole sample into $m$  Normal/non-central Chi-square distributed sub-groups;
	\item Applying the Johnson's transformation (only in the case of normally distributed sub-samples); 
	\item Calibrating the parameters of the CIR/Vasicek interest rate process $r$ to each sub-group by applying the optimal estimating function method described in Section \ref{sec:Calibration};
	\item Generating a sequence of estimated expected CIR/Vasicek interest rates by using the closed formula \eqref{Expectation} for each sub-group. 
\end{enumerate}

Again, we considered the above mentioned weekly observed data sample on derivatives with long-term maturity ($T$=30Y). From Step 2 we obtained a partition of the sample into $m=8$ normally distributed sub-groups (see Table~\ref{tab:RisulDatasetII1}), and into $m=42$ sub-groups with non-central Chi-square distribution (see Table~\ref{tab:RisulDatasetII2}). Note that the values $r_{308}$, in the first case, and $(r_{307},r_{308})$, in the second case,  were left out the partitioning. We then applied Steps 3-5 to each sub-group for both the partitions. 
Tables \ref{tab:RisulDatasetII1} and \ref{tab:RisulDatasetII2} list the RMSE computed for each sub-group, namely $\varepsilon_{k},$ with $k=1,...,m,$  
and the total RMSE, say $\widetilde{\varepsilon}$, computed over the whole sample as a weighted mean of the $\varepsilon_{k},$ that is
\begin{equation}\label{totalerror}
\widetilde{\varepsilon}= \sqrt{\sum_{k=1}^{m} \frac{n_k}{n}\sum_{h=1}^{n_k} e_h^{2}}.
\end{equation}    

\begin{table}[!ht] 
	\centering\caption{Partition by a Normal distribution:  error analysis of a sample of $n=308$ weekly EUR interest rates with maturity $T=$ 30Y from the Dataset II in Table \ref{tab:DatasetW}.}
		\scalebox{0.92}{
		\begin{tabular}{c|ccccc}
		\toprule
		& \multicolumn{5}{c}{\bf Normal distribution} \\
		\noalign{\smallskip}\hline\noalign{\smallskip}
		\textbf{Sub-group} & $(r_1,...,r_{33})$  & $(r_{34},...,r_{49})$ &  $(r_{50},...,r_{72})$ & $(r_{73},...,r_{202})$ & $(r_{203},...,r_{245})$ \\
		$\bm{\varepsilon_k}$  &0.1686  & 0.1181 & 0.0916 & 0.2538  & 0.2879\\
		\noalign{\smallskip}\hline\noalign{\smallskip}
		\textbf{Sub-group} & $(r_{246},...,r_{250})$ & $(r_{251},...,r_{267})$ & $(r_{268},...,r_{307})$ & &\\
	    $\bm{\varepsilon_k}$ & 0.0575 & 0.1182 & 0.1355 & &\\
		\noalign{\smallskip}\hline\noalign{\smallskip}
		$\bm{\widetilde{\varepsilon}}$ & \multicolumn{5}{c}{0.8663} \\ 
		\noalign{\smallskip}\hline
	\end{tabular}}
	\label{tab:RisulDatasetII1}
\end{table}

\begin{table}[!ht]
	\centering
	 \caption{Partition by a non-central Chi-square distribution: error analysis of a sample of $n=308$ weekly EUR interest rates with maturity $T=$ 30Y from Dataset II in Table \ref{tab:DatasetW}.}
	\scalebox{0.92}{
	\begin{tabular}{c|ccccc} 
	\toprule
	& \multicolumn{5}{c}{\bf Non-central Chi-square distribution} \\
	\noalign{\smallskip}\hline\noalign{\smallskip}
	\textbf{Sub-group} & $(r_1,...,r_{8})$  & $(r_{9},...,r_{16})$ &  $(r_{17},...,r_{23})$ & $(r_{24},...,r_{31})$ & $(r_{32},...,r_{40})$ \\
	$\bm{\varepsilon_k}$  &  0.0731 &  0.0701 &0.0372 & 0.1208 &0.1267\\
	\noalign{\smallskip}\hline\noalign{\smallskip}
	\textbf{Sub-group} & $(r_{41},...,r_{47})$ & $(r_{48},...,r_{55})$ & $(r_{56},...,r_{62})$ & $(r_{63},...,r_{69})$  & $(r_{70},...,r_{77})$\\
	$\bm{\varepsilon_k}$ &0.1079 &0.1024 &0.0396 &0.0675 &0.1241\\
	\noalign{\smallskip}\hline\noalign{\smallskip}
	\textbf{Sub-group}  &  $(r_{78},...,r_{84})$ & $(r_{85},...,r_{91})$ & $(r_{92},...,r_{98})$ & $(r_{99},...,r_{105})$ & $(r_{106},...,r_{112})$\\ 
	$\bm{\varepsilon_k}$ &0.0483 &0.1045 &0.0473 &0.0415 &0.0474\\
    \noalign{\smallskip}\hline\noalign{\smallskip}
    \textbf{Sub-group}  & $(r_{113},...,r_{119})$ & $(r_{120},...,r_{126})$  & $(r_{127},...,r_{133})$ &  $(r_{134},...,r_{140})$ & $(r_{141},...,r_{147})$\\
    $\bm{\varepsilon_k}$  &0.0671 &0.0586 &0.0579 &0.0782 &0.0439 \\
	\noalign{\smallskip}\hline\noalign{\smallskip}
	\textbf{Sub-group}  & $(r_{148},...,r_{154})$ & $(r_{155},...,r_{161})$ & $(r_{162},...,r_{168})$ & $(r_{169},...,r_{175})$ & $(r_{176},...,r_{182})$ \\ 
	$\bm{\varepsilon_k}$ &0.0311 &0.0557 &0.0250 &0.0312 &0.1061\\
    \noalign{\smallskip}\hline\noalign{\smallskip}
	\textbf{Sub-group}  & $(r_{183},...,r_{189})$ &  $(r_{190},...,r_{196})$ & $(r_{197},...,r_{203})$ & $(r_{204},...,r_{211})$ & $(r_{212},...,r_{218})$\\ 
	$\bm{\varepsilon_k}$ &0.0767 &0.0806 &0.0820 &0.0942 &0.0848\\
	\noalign{\smallskip}\hline\noalign{\smallskip}
	\textbf{Sub-group} & $(r_{219},...,r_{225})$ & $(r_{226},...,r_{235})$ & $(r_{236},...,r_{242})$  & $(r_{243},...,r_{249})$ &  $(r_{250},...,r_{256})$\\
	$\bm{\varepsilon_k}$ &0.0652 &0.1457 &0.0751 &0.1203 &0.0623\\
	\noalign{\smallskip}\hline\noalign{\smallskip}
	\textbf{Sub-group}  & $(r_{257},...,r_{263})$ & $(r_{264},...,r_{271})$ & $(r_{272},...,r_{278})$ & $(r_{279},...,r_{285})$ & $(r_{286},...,r_{292})$\\ 
	$\bm{\varepsilon_k}$  &0.0886 &0.0722 &0.0524 &0.0588 &0.0626\\
	\noalign{\smallskip}\hline\noalign{\smallskip}
	\textbf{Sub-group} & $(r_{293},...,r_{299})$ & $(r_{300},...,r_{306})$ & & & \\ 
	$\bm{\varepsilon_k}$ &0.0519 &0.0700 & & &\\
	\noalign{\smallskip}\hline\noalign{\smallskip}
	$\bm{\widetilde{\varepsilon}}$ & \multicolumn{5}{c}{0.0339}\\  
	\noalign{\smallskip}\hline
\end{tabular}}
\label{tab:RisulDatasetII2}
\end{table}

\begin{figure}[!htbp]
	\centering
	\includegraphics[width= 0.8\textwidth]{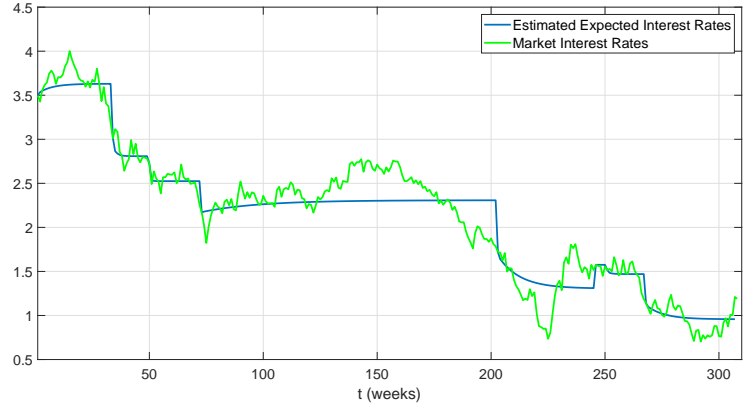}
	\caption{Estimated expected interest rates (blue line) versus market rates (green line) after segmentation with the Normal distribution for a data sample of $n=308$ weekly EUR interest rates with maturity $T=$ 30Y from Dataset II in Table \ref{tab:DatasetW}.}
	\label{fig:NormalDatasetII}
\end{figure}

\begin{figure}[!htbp]
	\centering
	\includegraphics[width=0.8\textwidth]{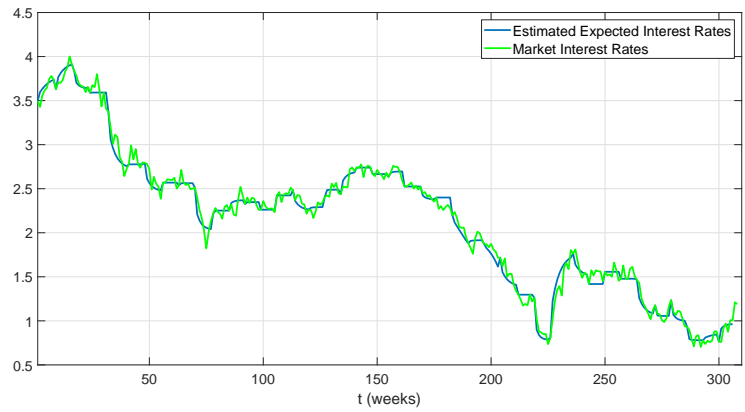}
	\caption{Estimated expected interest rates (blue line) versus market rates (green line) after segmentation with non-central Chi-square distribution for a data sample  of $n=308$ weekly EUR interest rates  with maturity $T=$ 30Y from Dataset II in Table \ref{tab:DatasetW}.}
	\label{fig:GammaDatasetII}
\end{figure}

Figures \ref{fig:NormalDatasetII} and \ref{fig:GammaDatasetII} below compare the plots of the estimated  expected interest rates with the original market data sample. From them is evident a better fitting to market data when we considered  partitioning data through non-central Chi-square distributed sub-samples.

\begin{figure}[!ht]
	\centering
	\includegraphics[width= 0.8\textwidth]{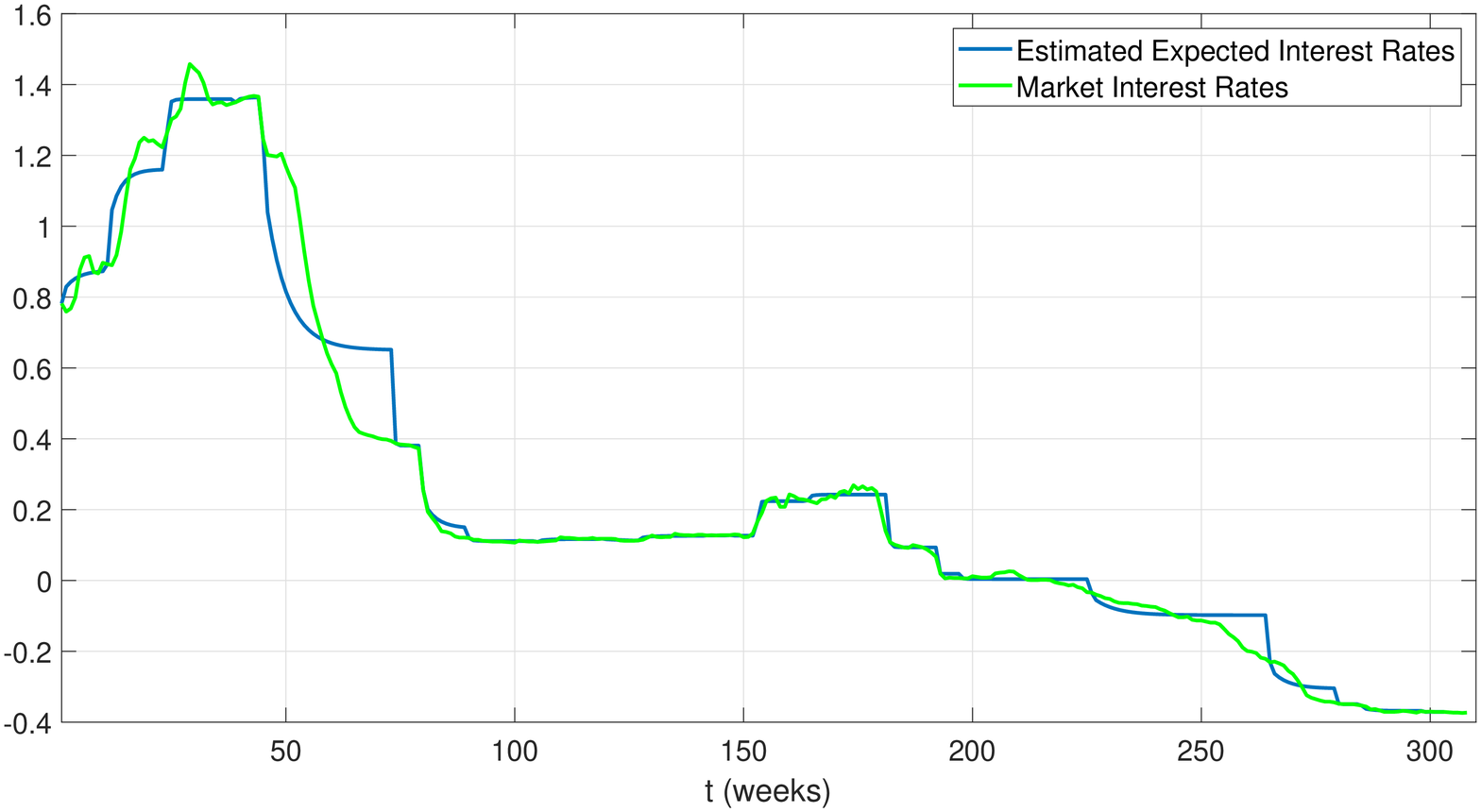}
	\caption{Estimated expected interest rates (blue line) versus market rates (green line) after segmentation with Normal distribution for a data sample  of $n=308$ weekly EUR interest rates  with maturity $T= 30/360A$ from Dataset I in Table \ref{tab:DatasetW}. }
	\label{fig:NormalDatasetI}
\end{figure}

\begin{figure}[!htbp]
	\centering
	\includegraphics[width=0.8\textwidth]{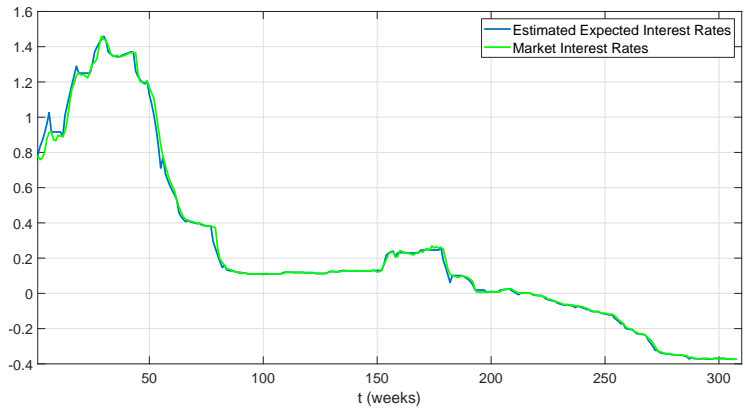}
	\caption{Estimated expected  interest rates (blue line) vs market rates (green line) after segmentation with non-central Chi-square distribution for a data sample  of $n=308$ weekly EUR interest rates  with maturity $T= 30/360A$ from Dataset I in Table \ref{tab:DatasetW}. }
	\label{fig:GammaDatasetI}
\end{figure}

\pagebreak
The second tested data sample consists of $n=308$  weekly EUR interest rates in a money market, on derivatives with maturity $T=$ 30/360A from Dataset I in Table \ref{tab:DatasetW}. From Step 2 of the above described numerical procedure, the entire sample has been partitioned into $m=23$ normally distributed sub-groups (see Table~\ref{tab:RisulDatasetI1}) and into $m=59$ non-central Chi-square distributed sub-groups (see Table~\ref{tab:RisulDatasetI2}). In this case, the observations $(r_{304}, r_{305}, r_{306}, r_{307}, r_{308})$ and $r_{308}$, respectively, were left out the partitioning. The results listed in Tables \ref{tab:RisulDatasetI1} and \ref{tab:RisulDatasetI2} as well as the plots in Figures \ref{fig:NormalDatasetI} and \ref{fig:GammaDatasetI}, show, also in this case, a better fitting to the observed money market interest rates when a partitioning into non-central Chi-square distributed sub-samples was considered. 

\begin{table}[!htbp]\centering
	\caption{Partition by a Normal distribution:  error analysis of a sample of $n=308$ weekly EUR interest rates with maturity $T= 30/360A$ from Dataset I in Table \ref{tab:DatasetW}.}
	\scalebox{0.92}{
\begin{tabular}{c|ccccc} 
		\toprule
		& \multicolumn{5}{c}{\bf Normal distribution} \\
		\noalign{\smallskip}\hline\noalign{\smallskip}
	\textbf{Sub-group} & $(r_1,...,r_{10})$  & $(r_{11},...,r_{23})$ &  $(r_{24},...,r_{38})$ & $(r_{39},...,r_{44})$ & $(r_{45},...,r_{73})$ \\
	$\bm{\varepsilon_k}$  &0.0435 &0.0933 &0.0470 & 0.0028 &0.2238\\\\
	\noalign{\smallskip}\hline\noalign{\smallskip}
	\textbf{Sub-group} & $(r_{74},...,r_{79})$ & $(r_{80},...,r_{89})$ & $(r_{90},...,r_{104})$ & $(r_{105},...,r_{118})$  & $(r_{119},...,r_{126})$\\
	$\bm{\varepsilon_k}$ &0.0038 &0.0220 &0.0019 &0.0029 &0.0016\\
	\noalign{\smallskip}\hline\noalign{\smallskip}
	\textbf{Sub-group}  & $(r_{127},...,r_{141})$ &  $(r_{142},...,r_{146})$ & $(r_{147},...,r_{152})$ & $(r_{153},...,r_{163})$ & $(r_{164},...,r_{181})$\\
	$\bm{\varepsilon_k}$  &0.0032 &0.0007 &0.0039 &0.0143 &0.0299\\
	\noalign{\smallskip}\hline\noalign{\smallskip}
	\textbf{Sub-group}  & $(r_{182},...,r_{192})$ &  $(r_{193},...,r_{197})$ & $(r_{198},...,r_{225})$ & $(r_{226},...,r_{264})$ & $(r_{265},...,r_{279})$\\ 
	$\bm{\varepsilon_k}$ &0.0097 &0.0107 &0.0143 &0.0488 &0.0315\\
	\noalign{\smallskip}\hline\noalign{\smallskip}
	\textbf{Sub-group}  & $(r_{280},...,r_{284})$ & $(r_{285},...,r_{298})$ & $(r_{299},...,r_{303})$ & &\\ 
	$\bm{\varepsilon_k}$ &0.0008 &0.0031 &0.0004 & &\\
	\noalign{\smallskip}\hline\noalign{\smallskip}
	$\bm{\widetilde{\varepsilon}}$ & \multicolumn{5}{c}{0.1700 } \\  
	\noalign{\smallskip}\hline
\end{tabular}}
	\label{tab:RisulDatasetI1}
\end{table}

\begin{table}[!htbp] \centering  \caption{Partition by a non-central Chi-square distribution: error analysis of a sample of $n=308$ weekly EUR interest rates with maturity $T= 30/360A$ from Dataset I in Table \ref{tab:DatasetW}.}
	\label{tab:RisulDatasetI2}
\small\addtolength{\tabcolsep}{-0.5pt}
\scalebox{0.96}{
	\begin{tabular}{c|ccccc} 
		\toprule
		& \multicolumn{5}{c}{\bf Non-central Chi-square distribution} \\
		\noalign{\smallskip}\hline\noalign{\smallskip}
		\textbf{Sub-group} & $(r_1,...,r_{6})$  & $(r_{7},...,r_{11})$ &  $(r_{12},...,r_{18})$ & $(r_{19},...,r_{24})$ & $(r_{25},...,r_{30})$ \\
		$\bm{\varepsilon_k}$  &0.0886 &0.0326 &0.0586 &0.0149 &0.0382\\
		\noalign{\smallskip}\hline\noalign{\smallskip}
		\textbf{Sub-group} & $(r_{31},...,r_{36})$ & $(r_{37},...,r_{42})$ & $(r_{43},...,r_{48})$ & $(r_{49},...,r_{55})$  & $(r_{56},...,r_{61})$\\
		$\bm{\varepsilon_k}$ &0.0148 &0.0046 &0.0435 &0.0854 &0.0306\\
		\noalign{\smallskip}\hline\noalign{\smallskip}
		\textbf{Sub-group}  &  $(r_{62},...,r_{66})$ & $(r_{67},...,r_{71})$ & $(r_{72},...,r_{76})$ & $(r_{77},...,r_{82})$ & $(r_{83},...,r_{87})$\\ 
		$\bm{\varepsilon_k}$ 	&0.0195 &0.0033 &0.0029 &0.0620 &0.0052\\
		\noalign{\smallskip}\hline\noalign{\smallskip}
		\textbf{Sub-group}  &  $(r_{88},...,r_{92})$ & $(r_{93},...,r_{97})$ & $(r_{98},...,r_{102})$ & $(r_{103},...,r_{107})$ & $(r_{108},...,r_{112})$\\ 
		$\bm{\varepsilon_k}$ &0.0021 &0.0003 &0.0021 &0.0006 & 0.0028\\
		\noalign{\smallskip}\hline\noalign{\smallskip}
		\textbf{Sub-group}  & $(r_{113},...,r_{117})$ & $(r_{118},...,r_{122})$  & $(r_{123},...,r_{127})$ &  $(r_{128},...,r_{132})$ & $(r_{133},...,r_{137})$\\
		$\bm{\varepsilon_k}$ & 0.0011 &0.0007 &0.0007 &0.0020 &0.0028\\
		\noalign{\smallskip}\hline\noalign{\smallskip}
		\textbf{Sub-group}  & $(r_{138},...,r_{142})$ & $(r_{143},...,r_{147})$ & $(r_{148},...,r_{152})$ & $(r_{153},...,r_{157})$ & $(r_{158},...,r_{162})$ \\ 
		$\bm{\varepsilon_k}$ &0.0008 &0.0004 &0.0049 &0.0122 &0.0115 \\
		\noalign{\smallskip}\hline\noalign{\smallskip}
		\textbf{Sub-group}  & $(r_{163},...,r_{167})$ & $(r_{168},...,r_{172})$ & $(r_{173},...,r_{177})$ & $(r_{178},...,r_{182})$ & $(r_{183},...,r_{187})$ \\ 
		$\bm{\varepsilon_k}$ & 0.0059 &0.0069 &0.0154 &0.0436 &0.0056 \\
		\noalign{\smallskip}\hline\noalign{\smallskip}
		\textbf{Sub-group}  & $(r_{188},...,r_{192})$ &  $(r_{193},...,r_{197})$ & $(r_{198},...,r_{202})$ & $(r_{203},...,r_{207})$ & $(r_{208},...,r_{212})$\\ 
		$\bm{\varepsilon_k}$ &0.0116 &0.0107 &0.0019 &0.0040 &0.0076\\
		\noalign{\smallskip}\hline\noalign{\smallskip}
		\textbf{Sub-group} & $(r_{213},...,r_{217})$ & $(r_{218},...,r_{222})$ & $(r_{223},...,r_{227})$  & $(r_{228},...,r_{232})$ &  $(r_{233},...,r_{237})$\\
		$\bm{\varepsilon_k}$ &0.0004 &0.0017 &0.0050 &0.0042 &0.0053\\
		\noalign{\smallskip}\hline\noalign{\smallskip}
		\textbf{Sub-group}  & $(r_{238},...,r_{242})$ & $(r_{243},...,r_{247})$ & $(r_{248},...,r_{252})$ & $(r_{253},...,r_{257})$ & $(r_{258},...,r_{262})$\\ 
		$\bm{\varepsilon_k}$  &0.0047 &0.0019 &0.0027 &0.0108 &0.0046\\
		\noalign{\smallskip}\hline\noalign{\smallskip}
		\textbf{Sub-group} & $(r_{263},...,r_{267})$ & $(r_{268},...,r_{272})$ & $(r_{273},...,r_{277})$ & $(r_{278},...,r_{282})$ & $(r_{283},...,r_{287})$ \\ 
		$\bm{\varepsilon_k}$ 	&0.0038 &0.0163 &0.0031 &0.0014 &0.0043\\
		\noalign{\smallskip}\hline\noalign{\smallskip}
		\textbf{Sub-group} & $(r_{288},...,r_{292})$ & $(r_{293},...,r_{297})$ & $(r_{298},...,r_{302})$ & $(r_{303},...,r_{307})$ &\\ 
		$\bm{\varepsilon_k}$  	&0.0013 &0.0020 &0.0016 &0.0006 &\\
		\noalign{\smallskip}\hline\noalign{\smallskip}
		$\bm{\widetilde{\varepsilon}}$ & \multicolumn{5}{c}{0.0323 }\\  
		\noalign{\smallskip}\hline
\end{tabular}}
\end{table}

\pagebreak
\subsection{Forecasting Expected Interest Rates}\label{sec:Forecast}
In this section we apply the proposed numerical procedure to forecast future expected values of market interest rates. We first consider a window of fixed size $m$ of market interest rates that is rolled through time, each time adding a new rate and taking off the oldest one. The length of this window  is the historical period over which we forecast the next expected spot rate value. It is worth noting that in case of large datasets, as with weekly observations, the methodology herein proposed necessarily requires to consider rolling windows of variable size. Thus, the step of the procedure relative to the dataset partition, described in Sections~\ref{sec:Partition}, is modified as follows. Fixed the initial length $m$ of the historical data sample, say $(r_{1+h},\dots,r_{m+h}), h=0,1,2,...$, the Lilliefors or the Kolmogorov-Smirnov test  is carried out starting from the last observed rate $r_{m+h}$ and then moving backward in the sample up to detect the smaller interest rate value $r_{m+h-s}\; (s=1,\dots, m-1)$, such that the sub-goup $(r_{m+h-s},\dots,r_{m+h})$ has Normal or non-central Chi-square distribution. 
The value $r_{m+h-s}$ denotes the \emph{change point} setting the window of latest interest rates, say $(r_{m+h-s},\dots,r_{m+h})$, which is taken into account to calibrate the parameters of the Vasicek/CIR model and forecast the next expected interest rate value $r_{m+h+1}$ (the past  interest rate values $(r_{1+h},\dots r_{m+h-s-1})$ are disregarded). Clearly, when the historical period is rolled through time, 
the size of the window over which to forecast a future rate may be variable since it depends on the detected change point.

\begin{figure}[!htbp]
	\centering
	\includegraphics[width=.8\textwidth]{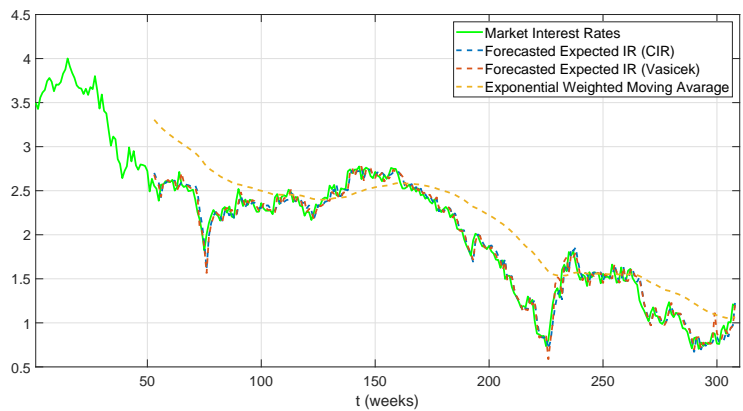}
	\caption{Forecast of expected next-week interest rates  based on a rolling window of variable size: sequence of $n=308$ weekly  EUR interest rates with maturity $T$=30Y (green line); CIR forecasted expected interest rates (blue dashed line); Vasicek forecasted expected interest rates (red dashed line); EWMA predicted values (yellow dashed line).}
	\label{Fig:for3}	
\end{figure}

We applied the modified numerical procedure to forecast expected future next-week interest rates. To explain our idea we refer to the first data sample considered in the previous section ($n=308$ weekly observed EUR interest rates with maturity $T=$ 30Y). The initial size of the historical data sample was fixed to $m=52$ weeks.  Steps 1-5 of the numerical procedure were applied to the historical market interest rates.  
Note that the calibration of the Vasicek/CIR model parameters with sub-groups of size smaller than 12 is not always possible when the optimal estimating function method mentioned in Section~\ref{sec:Calibration} is applied (see Remark \ref{Cal_remark}). In this case two adiacent sub-groups are joined together.
The sequence of forecast next-week expected values computed by formula \eqref{Expectation} for both Vasicek and CIR models, has been compared with the sequence of future rates computed by the Exponentially Weighted Moving Average (EWMA) by considering a rolling window of fixed size $m=52$ weeks.
The EWMA is a weighting scheme to estimate future values averaging on historical data with weights that decrease exponentially at a rate $\lambda$ throughout as the observations are far in the past (the reader can refer, for example, to Hull (Hull, 2012, Ch.II). 
The EWMA has been shown to be powerful for prediction over a short horizon and track closely the volatility as it changes. Indeed recent interest rates movement is the best predictor of future movement as it is not conditioned on a mean level of volatility. 
The forecasted sequences are plotted in Figure \ref{Fig:for3}. The proposed numerical procedure clearly shows a better performance with respect to the EWMA model. 

\begin{figure}[!htbp]
	\centering
	\includegraphics[width=0.8\textwidth]{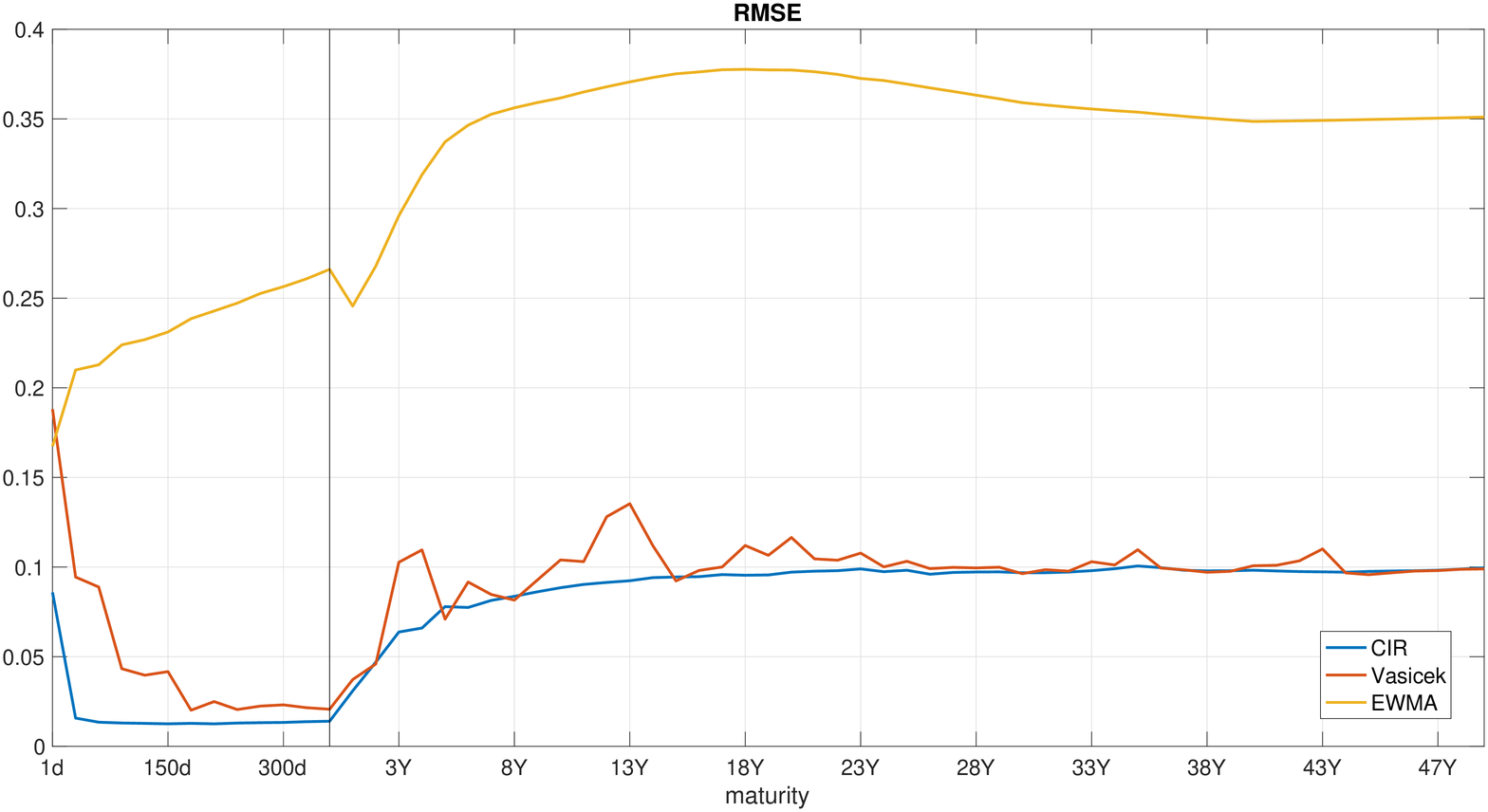}
	\caption{Error analysis in the forecast procedure of future next-week expected interest rates. RMSE values computed by the proposed numerical procedure based on a rolling window of variable size: CIR model (blu line), Vasicek model (red line). RMSE values computed by the EWMA model (yellow line) based on a rolling window of fixed size $m=52$ weeks. The vertical black line differentiates samples of Dataset I from samples of Dataset II in Table~\ref{tab:DatasetW}.}
	\label{Fig:RMSEvariable}	
\end{figure}

Further, an error analysis to all data samples (63 maturities) available from Dataset I and II  in Table~\ref{tab:DatasetW} was carried out.  Figure \ref{Fig:RMSEvariable} compares the corresponding RMSE values computed by applying our numerical algorithm  to the Vasicek and CIR model with the ones computed by the EWMA model. The initial size of the historical data sample was fixed to $m=52$ weeks (the vertical black line differentiates samples of Dataset I from samples of Dataset II). In this case, too, a better performance of our procedure with respect to the EWMA model is confirmed. Further, the modified procedure  to make predictions based on rolling windows of variable size highlights an improvement in forecast when a partition with non-central Chi-square distribution (CIR model) is considered.

\section{Conclusions}\label{sec:Conclusions}
In this paper, we have presented a new methodology for using both the Vasicek and the CIR model in order to forecast interest rates that works even when interest rates are negative. To achieve that objective, we have proposed a numerical procedure  partitioning the selected data sample according to the best fitting of  a Normal or non-central Chi-square  distribution. These distributions were chosen by analogy with the steady (resp. conditional) distribution of the interest rate process in the Vasicek (resp. CIR) model. Where the first was taken when the steady distribution of the interest rate process is modelled with Vasicek and the second when the conditional distribution of the said process is represented by the CIR model.
After having partitioned the sample of observed market data, the Vasicek/CIR model's parameters are calibrated to each sub-sample of market spot rates and the corresponding expected interest rates are estimated by the conditional expectation closed formula of both models. We have also included in the procedure a step concerning an appropriate translation of market interest rates to positive values in order to overcome the issue of negative/near-to-zero values which are not compatible with the CIR model. Finally,  we have analyzed the empirical performance of the proposed methodology for two different weekly recorded EUR data samples in a money market and a long-term dataset, respectively.   Better results are shown in terms of the RMSE  when a segmentation of the data sample in non-central Chi-square distributed sub-samples is considered. After assessing the accuracy of our procedure, we have applied the implemented algorithm to forecast future expected interest rates over rolling windows of historical data with  variable size. 
The performance of the new approach, tested on weekly rates with different maturities, has been carried out for both Vasicek and CIR model, and compared with the EWMA model  in terms of forecasting error. The error analysis highlighted a better performance of the proposed procedure with respect to EWMA model and better results in prediction when a partition of the historical data sample in non-central Chi-square distributed sub-samples (CIR model) is considered.

%
%



\end{document}